%% file: main.tex
\documentclass[sigconf]{acmart}
\usepackage{etc}
\usepackage{float}
\usepackage{subcaption}
\usepackage{cleveref}
\usepackage{tcolorbox}
\usepackage{nicematrix, tikz}
\usepackage{subcaption}
\usepackage{bbm}
\usepackage{multicol}
\usetikzlibrary{fit}
\AtBeginDocument{%
  \providecommand\BibTeX{{%
    \normalfont B\kern-0.5em{\scshape i\kern-0.25em b}\kern-0.8em\TeX}}}

\author{Mootez Saad and Tushar Sharma}
\email{{mootez, tushar}@dal.ca}
\affiliation{%
  \institution{Faculty of Computer Science}
  \city{Halifax}
  \state{Nova Scotia}
  \country{Canada}
}

\begin{document}

\title{Naturalness of Attention:\\ Revisiting Attention in Code Language Models}

\acmConference[ICSE 2024]{46th International Conference on Software Engineering}{April 2024}{Lisbon, Portugal}

\setcopyright{none}

\begin{abstract}
Language models for code such as CodeBERT offer the capability to learn advanced source code representation, but their opacity poses barriers to understanding of captured properties. 
Recent attention analysis studies provide initial interpretability insights by focusing solely on attention weights rather than considering the wider context modeling of Transformers. 
This study aims to shed some light on the previously ignored factors of the attention mechanism beyond the attention weights.
We conduct an initial empirical study analyzing both attention distributions and transformed representations in CodeBERT. Across two programming languages, Java and Python, we find that the scaled transformation norms of the input better capture syntactic structure compared to attention weights alone. Our analysis reveals characterization of how CodeBERT embeds syntactic code properties. The findings demonstrate the importance of incorporating factors beyond just attention weights for rigorously understanding neural code models. This lays the groundwork for developing more interpretable models and effective uses of attention mechanisms in program analysis.
\end{abstract}




\maketitle

\section{Introduction}
\input{section/1_introduction}
\section{Background and Motivation}
\label{background}
\input{section/2_background}
\section{Experiments and Results}
\input{section/3_methodology}
\section{Conclusions and future work}
\input{section/4_future_plans}

\onecolumn
\begin{multicols}{2}
\bibliographystyle{ACM-Reference-Format}
\bibliography{sample-base}
\end{multicols}

\end{document}

%% file: section/1_introduction.tex
Obtaining effective representations of source code is crucial for many program analysis tasks such as 
code search, 
code completion, 
and program translation. 
Earlier code representation approaches, such as Code2Vec~\cite{Alon2019} and Code2Seq~\cite{Alon2018codeseq} 
demonstrated initial progress in learning distributed vector representations of code.
However, these methods are limited in their modeling capacity 
and do not fully capture the rich semantics of code.
To overcome the limitations in early code representation techniques,
Transformer-based~\cite{Vaswani2017} neural models have emerged as a promising paradigm for learning effective code representations. 
Inspired by their success in natural language processing, architectures such as BERT~\cite{Devlin2019} have been adapted to code representation,
leading to the emergence of Language Models of Code (LMC), such as CodeBERT~\cite{Feng2020}, GraphCodeBERT~\cite{Guo2021}, CodeT5~\cite{Wang2021}, and Code Llama~\cite{Rozière2023}. 
The self-attention mechanism powering Transformers provides stronger representational capabilities compared to earlier architectures such as Recurrent Neural Networks (RNNs). 
This has enabled Transformer-based models to establish new state-of-the-art results across a variety of software engineering tasks that involve source code analysis, processing, and manipulation by learning better semantic representations of programs.

Despite such advances, a major limitation of LMCs is their black box nature and lack of interpretability.
While models such as CodeBERT show impressive performance on downstream tasks,
it is unclear which properties of code they capture or learn internally.

To address these interpretability issues, an emerging area of research involves analyzing and probing these complex neural networks through attention visualization and representation analysis. 
Sharma \etal{}~\cite{Sharma2022} found BERT models trained on code exhibit key attention differences from natural language - namely, higher focus on identifiers over special tokens like \texttt{[CLS]} and more localized context.
Wan \etal{}~\cite{Wan2022} investigated the encoded syntactic patterns within the attention distributions of CodeBERT and GraphCodeBERT.
However, in the realm of natural language, Kobayashi \etal{}~\cite{Kobayashi2020} note that attention weights alone may not reveal the full perspective of the patterns learned by the model. Based on the \textit{naturalness} property of software~\cite{Hindle2021}, techniques effective for analyzing natural language models may also lend insight into source code models.
This motivates revisiting attention-based analysis by incorporating the factors that were previously ignored when analyzing LMCs.

In this paper, we revisit the mathematical formulation of the Multihead Attention (\mhattn{})~\cite{Kobayashi2020} to illustrate how it is composed of two factors: \textit{attention weights} and the \textit{transformation of input}. 
Given this new reformulation, we perform a trend analysis of the Transformer layers of CodeBERT on Java and Python to study the differences between these two factors.
We show how including the previously ignored effect leads to a better alignment with the syntactic properties of source code compared to the attention weights.

We make the following contributions:
\begin{itemize}
\item 
First, we perform a layer-wise analysis to study the trends of attention weights \textbf{and} the transformation of input for an LMC (\ie{} CodeBERT).
To the best of our knowledge, this is the first study that emphasizes the need to consider the transformation of input along with attention weights in the context of an LMC.
\item 
Second, we compare the capacity of the two factors to capture the syntactic properties of source code.

\end{itemize}
We make the replication package, including code and data, available online~\cite{replication}.

%% file: section/2_background.tex
The core component of the Transformer architecture is the Multiheaded Attention (\mhattn{}), which is composed of multiple Self-Attention (SA) heads. For instance, in CodeBERT, which is based on RoBERTa's~\cite{Liu2019} architecture, each \mhattn{} layer is composed of 12 SA heads. Let $h$ denote the number of heads, and $X$ be the input sequence of length $n$, where each token is embedded in $\mathbb{R}^{d'}$ ($d'=64$ in CodeBERT). 
In each head, a sequence is projected into three matrices: \qmat{} ($Q_{n\times d'}$), \kmat{} ($K_{n\times d'}$) and \vmat{} ($V_{n\times d'}$). Formally, these matrices are defined as follows:
\newline
For $i \in [1\ldots h]$
\begin{equation}
    Q^{(i)} = X\cdot W_{Q}^{(i)}, \; K^{(i)} = X\cdot W_{K}^{(i)}, \; V^{(i)} = X\cdot W_{V}^{(i)}
\label{eqn:qkv}
\end{equation}

The attention matrix $\Alpha$ is computed by applying the \texttt{softmax(.)} function on the result of the multiplication of the \qmat{} and \kmat{} matrices\footnote{In Vaswani \etal{}~\cite{Vaswani2017}, this multiplication includes an optional mask to mask out certain tokens such as padding tokens. We omit this for the sake of simplicity.}, scaled by the square root of their dimension $d'$.
\begin{equation}
    \Alpha^{(i)} = softmax(\frac{ Q^{(i)} {K^{(i)}}^{T}}{\sqrt{d'}})
\label{eqn: attn_matrix}
\end{equation}
Then, the attention matrix is multiplied by the \vmat{} matrix to obtain the attention output $z$,

\begin{equation}
    z^{(i)} = \Alpha^{(i)}\cdot V^{(i)}
\label{eqn: attn_otpt}
\end{equation}

Finally,
concatenating the output of each head and multiplying it by a weight matrix $W^O_{hd'\times hd'}$ gives the output of the \mhattn{} layer.
\begin{align}
Z_{\text{MHA}} &= [z_{1}; \ldots ; z_{h}]_{n\times hd'}, \\
Y_{\text{MHA}} &= Z\cdot W^{O} 
\label{eqn: mha}
\end{align}

$Y_{\text{MHA}}$ can be reformulated given the linearity of matrix multiplication. To build the intuition, let us consider the calculation of the entry located at the 1\textsuperscript{st}-row and 1\textsuperscript{st}-column of $Y_{\text{MHA}}$. It is done by taking the dot product of the 1\textsuperscript{st}-row of $Z_{\text{MHA}}$ and the 1\textsuperscript{st}-column of $W^{O}$.
\begin{equation}
    \tag{6}
    Y_{\text{MHA}}[1, 1] = \sum_{i=1}^{hd'}{Z_{\text{MHA}}[1,i] W_{O}[i,1]}
\label{eqn: ymha_11}
\end{equation}
We can decompose~\Cref{eqn: ymha_11} into $h$ summations,
\begin{align}
     \begin{split}
        \tikz \node[] {$Y_{\text{MHA}}[1, 1]=$}; 
        & \tikz \node[opacity=.2, text opacity=1, fill=red!25,rounded corners,inner sep=3pt] {$\sum_{i=1}^{d'}{Z_{\text{MHA}}[1,i] W_{O}[i,1]}$}; \\
        \tikz \node[] {$+$};& \tikz \node[opacity=.2, text opacity=1, fill=green!25, rounded corners, inner sep=3pt] {$\ldots$}; \\
        \tikz \node[] {$+$};& \tikz \node[opacity=.2, text opacity=1, fill=blue!25, rounded corners, inner sep=3pt] {$\sum_{i=hd'-d'+1}^{hd'}{Z_{\text{MHA}}[1,i] W_{O}[i,1]}$};
     \end{split}
\tag{6}
\end{align}
\vspace{-4.5mm}

\input{figures/z_mha_wo_mult_2_col}
\vspace{-3mm}
By extension and with reference to~\Cref{fig:zmha_wo}, we can express $Y_{\text{MHA}}$ as the sum of $h$ matrices calculated from the multiplication of the submatrices from $Z_{\text{MHA}}$ and $W_{O}$. This entails that \Cref{eqn: mha} can be rewritten as follow,
\begin{equation}
\tag{9}
    Y_{\text{MHA}} = Z_{\text{MHA}}\cdot W^{O} = \sum_{i=1}^{h}{z^{(i)}\cdot W_{O}^{(i)}}
\label{eqn: new_ymha}
\end{equation}
and if we plug in \Cref{eqn: attn_otpt} and \Cref{eqn:qkv} we obtain,
\begin{align}
\label{eqn:eqlabel}
\tag{10}
\begin{split}
    Y_{\text{MHA}} &= \sum_{i=1}^{H}{\Alpha^{(i)}\cdot V^{(i)}\cdot W_{O}^{(i)}}
    \\
    &= \sum_{i=1}^{H}{\Alpha^{(i)}X\cdot W_{V}^{(i)}\cdot W_{O}^{(i)}} 
    \\
    &= \sum_{i=1}^{H}{\Alpha^{(i)} f^{(i)}(X)}
\end{split}
\end{align}
We can see from this reformulation that the information $Y_\text{MHA}$ holds is the result of the contribution from \textit{two} factors: the attention weights $\Alpha$ \textit{and} the transformation $f(.)$ applied on the input $X$. 
A token $t_{i} \in X$ can have a high attention weight $\alpha_{i} > 0$ and, at the same time, a low contribution from its transformation $||f(t_{i})|| \approx 0$~\cite{Kobayashi2020}. \textit{This implies that the properties deduced from the analysis of \mhattn{} cannot solely be attributed to the attention weights}. The previous studies that performed attention analysis of Language Models of Code (LMC) concentrated on probing the attention weights to see the type of patterns they exhibit and how well they align with the properties of source code~\cite{Wan2022, Sharma2022}. 

Motivated by this argument, we would like to revisit the attention analysis of LMC by considering the missing factor, \ie{} the scaled transformation $||\alpha f(x)||$. 

The primary objective is to thoroughly comprehend the attention mechanism's properties when applied to source code.
To this end, we conduct a preliminary empirical study and answer the following research questions:
\begin{itemize}
    \item[\textbf{RQ1:}] How do the general trends across layers between attention weights \attn{} and the scaled transformations norms \afx{} compare?
    \item[\textbf{RQ2:}] How does \afx{} align with the syntactic structure of source code compared to attention weights?
\end{itemize}

%% file: figures/z_mha_wo_mult_2_col.tex
\begin{figure}[H]
    \begin{subfigure}[t]{.4\columnwidth}
    \hspace{-8mm}
        \resizebox{1.2\columnwidth}{!}{%
        $\displaystyle
        {\scalebox{1.8}{$Z_{\text{MHA}}$}} {\scalebox{1.7}{$=$}}
        \begin{bNiceMatrix}[right-margin = 6pt, left-margin = 6pt] 
        \text{z}_1[1,1] & \ldots & \text{z}_1[1,d'] & \ldots &\text{z}_h[1,1] & \ldots & \text{z}_h[1,d'] \\  
        \text{z}_1[2,1] & \ldots & \text{z}_1[2,d'] & \ldots & \text{z}_z[2,1] & \ldots & \text{z}_z[2,d'] \\ 
        \vdots & \vdots & \vdots & \vdots &  \vdots & \ldots & \vdots \\
        \text{z}_1[n,1] & \ldots & \text{z}_1[n,d'] & \ldots & \text{z}_z[n,1] & \ldots & \text{z}_z[n,d'] \\ 
        \CodeAfter
            \tikz \node [fill=red!25, opacity=.2, draw, rounded corners, fit = (1-1) (last-3)] {} ; 
            \tikz \node [fill=green!25, opacity=.2, draw, rounded corners, fit = (1-4) (last-4)] { } ;
            \tikz \node [fill=blue!25, opacity=.2, draw, rounded corners, fit = (1-5) (last-7)] { } ;
          \OverBrace[shorten,yshift=3mm]{1-1}{last-3}{{\scalebox{1.7}{$\mathrm{z^{(1)}_{n\times d'}}$}}}
          \OverBrace[yshift=3mm]{1-4}{1-4}{{\scalebox{1.7}{$\ldots$}}}
          \OverBrace[shorten,yshift=3mm]{1-5}{1-7}{{\scalebox{1.7}{$\mathrm{z^{(h)}_{n\times d'}}$}}}
        \end{bNiceMatrix}_{{\scalebox{1.3}{$n \times hd'$}}}
        $
        }

\end{subfigure}
\hspace{-3mm}
\begin{subfigure}[t]{.5\columnwidth}
\resizebox{1.1\columnwidth}{!}{%
$\displaystyle
{\scalebox{1.4}{$W_{O}$}} =
\begin{bNiceMatrix}[right-margin = 10pt, left-margin = 6pt] 
\text{w}[1,1] & \ldots  & \text{w}[1,hd'] \\
\vdots & \vdots  & \vdots  \\
\text{w}[d',1] &  \ldots & \text{w}[d',hd'] \\
 &  &      \\
\vdots & \vdots & \vdots  \\
 &  &    \\
\text{w}[hd'{-}d',1] & \ldots  & \text{w}[hd'{-}d',hd'] \\
\vdots & \vdots & \vdots  \\
\text{w}[hd',1] & \ldots & \text{w}[hd',hd'] \\
\CodeAfter
\SubMatrix{\{}{1-1}{3-2}{.}[left-xshift=15pt, name=A] {}
\tikz \node [left] at (A-left.west) {$W_{O}^{(1)}$} ;
  \tikz \node [fill=red!25, opacity=.2, draw, rounded corners, fit = (1-1) (3-last.south east)] {} ; 
  \tikz \node [fill=green!25, opacity=.2, draw, rounded corners, fit = (5-1) (5-last)] { } ;
  \tikz \node [shift={(0.2,0)}, fill=blue!25, inner sep=8pt, opacity=.2, draw, rounded corners, fit = (7-1) (9-last)] { } ;
\end{bNiceMatrix}_{hd' \times hd'}
$
}
\end{subfigure}
\caption{
Multiplication of the concatenated outputs of each SA head with the weight matrix. The highlighted blocks are the components that are attributed to each SA head.
}
\label{fig:zmha_wo}
\end{figure}

%% file: section/3_methodology.tex
\input{figures/alpha_afx_trends}
In this section, we will present our methodology and the conducted experiments to answer the research questions stated above. 

\subsection{Methodology}
\textbf{Models}: In this study, we considered CodeBERT~\cite{Feng2020}, a pretrained language model of code that adopts the architecture and pertaining strategy as RoBERTa~\cite{Liu2019}. We chose such a model to follow Wan \etal{}~\cite{Wan2022}. In addition, it is one of the earliest LMCs which has spawned many follow-up works. Analyzing it provides a strong baseline for future comparative studies on other related models. 

It consists of $12$ Transformer~\cite{Vaswani2017} layers, each encompassing $12$ self-attention heads. It was trained on a set of bimodal instances (\ie{} pairs of natural language and programming languages), across six programming languages from the CodeSearchNet dataset~\cite{Husain2019}. 

\noindent

\textbf{Data:} We used CodeSearchNet dataset~\cite{Husain2019} to create corpora for two programming languages: Java and Python. 
Each corpus consists of $5,000$ randomly sampled code snippets with lengths less than $512$ tokens. 

\vspace{-2mm}
\subsection{RQ1---Trend Analysis}

Through this research question, we aim to investigate 
the trends, at a macro level, in the behaviours of the attention weights $\alpha$ versus the scaled norms \afx{} across the layers of CodeBERT.

\noindent
\textbf{Approach:} 
From each self-attention head, we extracted the attention weights \attn{}, transformation norm $||f(x)||$ and scaled transformation norm \afx{} matrices, for each instance. Since CodeBERT was trained using a \textit{WordPiece} tokenizer~\cite{Wu2016},  each word can be tokenized further into subtokens. Given that our analysis was carried out at a word level, we follow the same procedure done by Clark \etal{}~\cite{Clark2019} and convert token-token maps to word-word maps by taking the average of a word's subtokens. To make our analysis granular, we group tokens by their categories: \textit{Keywords}, \textit{Identifiers}, \textit{Literals}, \textit{Operators} and \textit{Special Symbols}. Each category is defined according to each programming language grammar specification (Java~\cite{Oracle2023} and Python~\cite{PSF2023}).

\noindent
\textbf{Results}: 
\Cref{fig:trend_alpha_afx_main} depicts the variation of the average attention weights and the average scaled transformation norms across each layer for each token category in both datasets. 
 Aligned with the findings of Kobayashi \etal{}~\cite{Kobayashi2020} and Clark \etal{}~\cite{Clark2019}, and in contrast to the results of Sharma \etal{}~\cite{Sharma2022}, the special tokens displayed higher average attention compared to other token types. Specifically, \clstok{} had the highest average attention between Layers 2 and 4, which then decreased until Layer 7. Then, its attributed attention kept on increasing until Layer 10. Interestingly, the drop in \clstok{}'s attention between Layers 5 and 7 appeared to transfer to \septok{}, whose average attention peaked within this range. This pattern held across programming languages, with similar trends in both Java and Python corpora, though minor differences arose in precise attention values. For instance, \clstok{} attention remained constant between Layers 2 and 3 for Java, whereas it slightly declined in the Python dataset. 

However, this pattern is different when calculating the \textit{scaled transformation norms} \afx{}. The contribution of these tokens is lower compared to other categories such as \textit{Identifiers} and \textit{Special Symbols}. 
This indicates, that similar to BERT,
when CodeBERT does not find information in the input,
it assigns higher attention values to these tokens given that the attention weights should sum up to 1 (due to the \texttt{softmax(.)} function).

Although the \textit{ranking} of each token category at each layer appears to be consistent between \attn{} and \afx{}, there is some contrast between the patterns observed at each layer. For example, if we look at the trend of attention weights of the \textit{Keyword} tokens in the Python dataset in~\Cref{fig:trend_alpha_afx_python}, we see that the attention values drop between Layer 1 and Layer 2 and remain relatively constant between Layers 2 and 4. In contrast, the values of \afx{} between Layers 1 and 3 are increasing for this category. This contrast effect is also observed for other types such as \textit{Identifiers}. Generally, in both datasets, we see that in some layers, when the attention weights are constant (\eg{} L2-L4 and L5-L8) the scaled transformation norms exhibit either a peak or a decline. One explanation can be the \textit{cancelling} effect of \attn{} and $||f(x)||$ that was mentioned in Section~\ref{background} 
, hence, the contrast between \attn{} and \afx{}. \Cref{fig:alpha_vs_fx} further illustrates this cancelling effect for the special tokens \clstok{} and \septok{}, and the \textit{Literals} category. 
\vspace{-2mm}
\smrybx{\textbf{Summary}: The results show that the behavior of attention weights and the scaled transformation norms \afx{} differ significantly. 
The two components of Multiheaded Attention \ie{} \attn{} and $||f(x)||$, often exhibit a cancellation effect.
Such contrast entails that including other variables, \ie{} the transformation norm $||f(x)||$, when performing attention analysis, might lead us to more comprehensive and explainable results. 
In other words, extending the analysis to regions other than attention weights might reveal additional insights about the language model's capacity to model the relations and properties of source code.}
\vspace{-4mm}

\input{figures/spec_tokes_ltrl_maps}
\subsection{RQ2---Syntactic Alignment}
In this section, we analyze the syntactic properties that are embedded in \afx{} compared to those in the attention weights. 

\noindent
\textbf{Approach}: \citet{Vig2021} proposed a metric, $p_{\alpha}(g)$, that measures the agreement between an attention map (\ie{} the attention matrix or weights) and a property map generated by an indicator function $g$. The function $g(i, j)$ returns 1 if a given property exists between two tokens $i$ and $j$, 0 otherwise. Wang \etal{}~\cite{Wan2022} defined $g$ to return 1 if the pair ($i$, $j$) share the same parent in the Abstract Syntax Tree (\abst{}) of a code snippet $x$. Their intuition was that attention defines the closeness of each pair of code tokens. This score is formally defined in~\Cref{eqn:syntct_agrmnt},
\begin{equation}
\tag{11}
{p_\alpha}(g) = \frac{{  \sum\limits_{\mathbf{x} \in \mathbf{X}} \sum\limits_{i=1}^{|\mathbf{x}|}\sum\limits_{j=1}^{|\mathbf{x}|}
    {f(i, j)} \cdot \mathbbm{1}_{\alpha_{i,j} > \theta} 
  }}{{ \sum\limits_{\mathbf{x} \in \mathbf{X}}\sum\limits_{i=1}^{|\mathbf{x}|}\sum\limits_{j=1}^{|\mathbf{x}|}
    \mathbbm{1}_{\alpha_{i,j} > \theta} }}
\label{eqn:syntct_agrmnt}
\end{equation}
where $\mathbbm{1}_{\alpha_{i,j} > \theta}$ is an indicator function that selects high-confidence attention weights ($\theta = 0.3$ in~\cite{Vig2021, Wan2022}). In other words, $\mathbbm{1}_{\alpha_{i,j} > \theta}$ evaluates to 1 if $\alpha_{i,j} > \theta$, and 0 otherwise.
\Cref{eqn:syntct_agrmnt} sums over all token pairs ($i$, $j$) where the attention $\alpha_{i,j} > \theta$. It counts how many of these high-confidence attention pairs connect tokens that are syntactically related according to the AST (\ie{} $f(i, j) = 1$) over the dataset.
Dividing this count by the total number of high-confidence pairs gives the proportion of attention connections that align with the AST structure. This proportion indicates how well the attention matches syntactic relationships. 
\vspace{-2mm}
\input{figures/ast_alpha_afx_agreement}
\vspace{-4mm}
\noindent
\textbf{Results}: 
\Cref{fig:ast_agreement} illustrates the degree each head in each of CodeBERT's layers aligns with the code's syntactic properties by considering attention weights and \afx{}. Overall, it appears that the scaled transformation embeds better the syntactic properties of source code than the bare attention weights. Interestingly, it even captures such properties starting at earlier layers. For instance, we see that in Layer 2, the maximum agreement percentage is 83.4\% compared to 54.8\% in the attention maps, and it is $\times2.6$ more aligned in the first layer. The same trend is also observed in the majority of the remaining layers, whether in the middle (\eg{} Layer 4, 76.7\% vs 38.8\% and Layer 5, 63.9\% vs 34.7\%) or at the end (Layer 10, 59.5\% vs 57\% and Layer 12, 48.1\% vs 37.4\%).
However, there are layers where the attention weights exhibit a higher alignment than \afx{}. Notably, in Layers 3, 7 and 10 (34\% vs 27.9\%, 27\% vs 26.1\%, and 67.8\% vs 66.6\%).
However, the alignment with attention weights is significantly lower than the alignment with scaled transformation

\smrybx{\textbf{Summary}: Generally, the scaled transformation norms show a higher alignment with the syntactic properties of source code. However, there are regions where the attention weights model better such properties. These observations contribute to our understanding of how the attention mechanism embeds implicit code patterns.}

%% file: figures/alpha_afx_trends.tex
\begin{figure*}
 
    \begin{subfigure}[b]{\columnwidth}
        \includegraphics[width=\columnwidth]{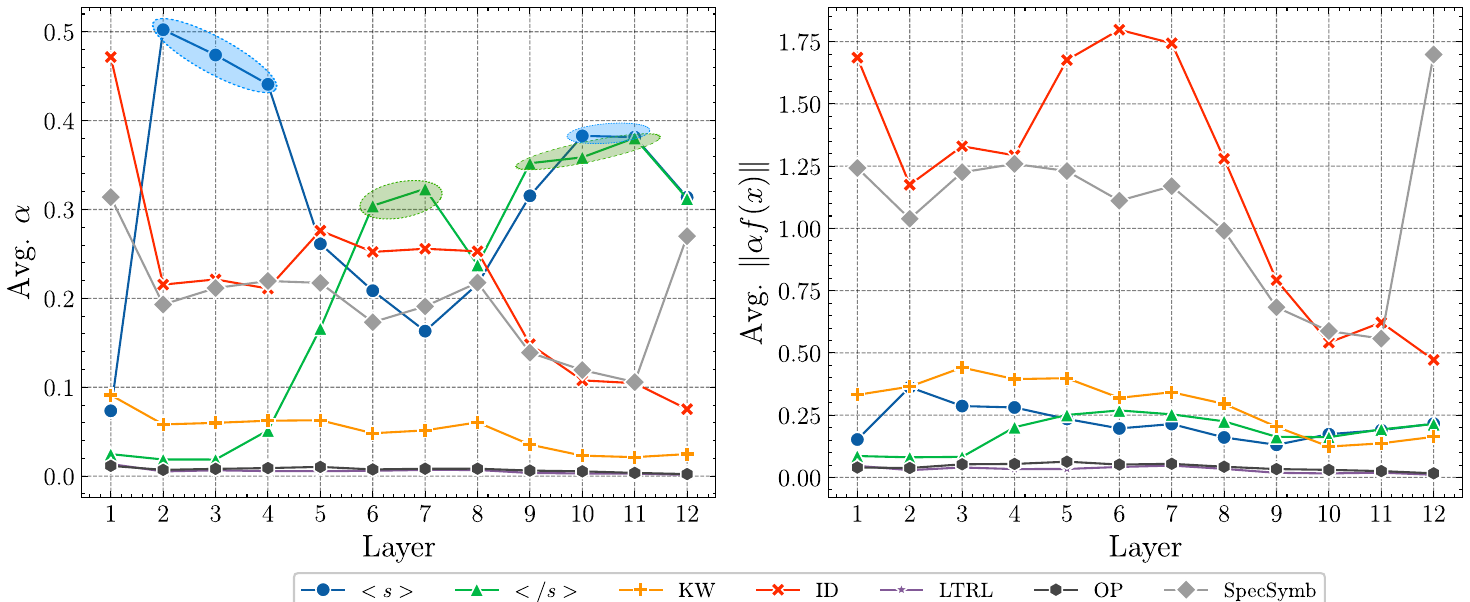}
        \caption{Python}
        \label{fig:trend_alpha_afx_python}
        \end{subfigure}
    ~
    \begin{subfigure}[b]{\columnwidth}
        \includegraphics[width=\columnwidth]{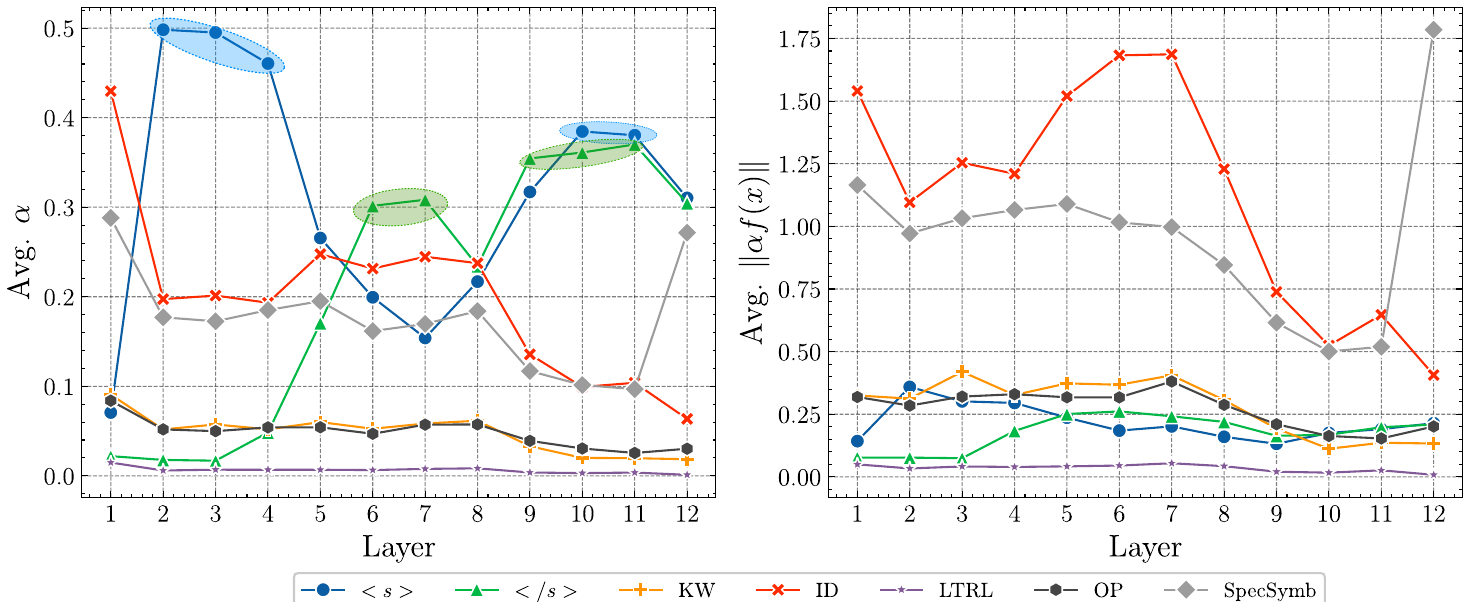}
        \caption{Java}
        \label{fig:trend_alpha_afx_java}
        \end{subfigure}
    \caption{Variation of the average \attn{} and \afx{} values of each token category across layers in CodeBERT. Abbreviations: Keywords (\texttt{KW}), Identifiers (\texttt{ID}), Literals (\texttt{LTRL}), Operators (\texttt{OP}), Special Symbols (\texttt{SpecSymb}). Note that \texttt{<s>} and \texttt{</s>} are the classification and separation tokens respectively. In other BERT variants, they are represented by \texttt{[CLS]} and \texttt{[SEP]}.
  }
    \label{fig:trend_alpha_afx_main}
\end{figure*}


%% file: figures/spec_tokes_ltrl_maps.tex
\begin{figure*}
    \begin{subfigure}[b]{.33\textwidth}
    \hspace{-2cm}
        \begin{subfigure}[b]{.15\textwidth}
        \includegraphics[scale=.26]{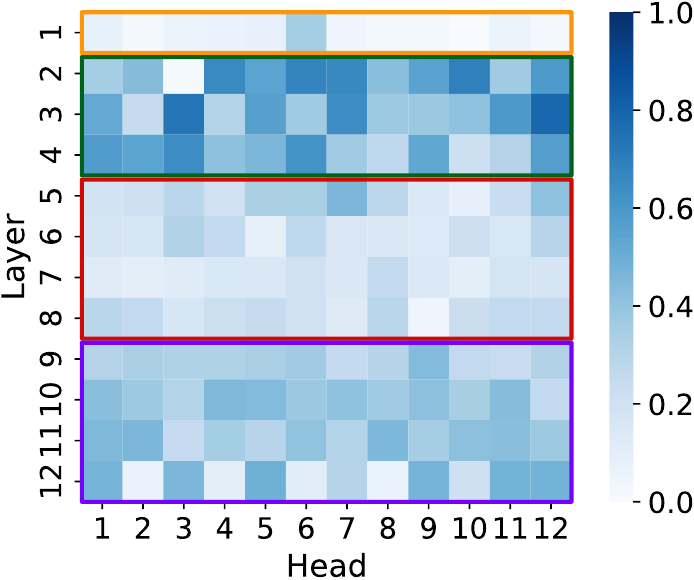}
        \end{subfigure}
        \hspace{2.2cm}
        \begin{subfigure}[b]{.15\textwidth}
        \includegraphics[scale=.26]{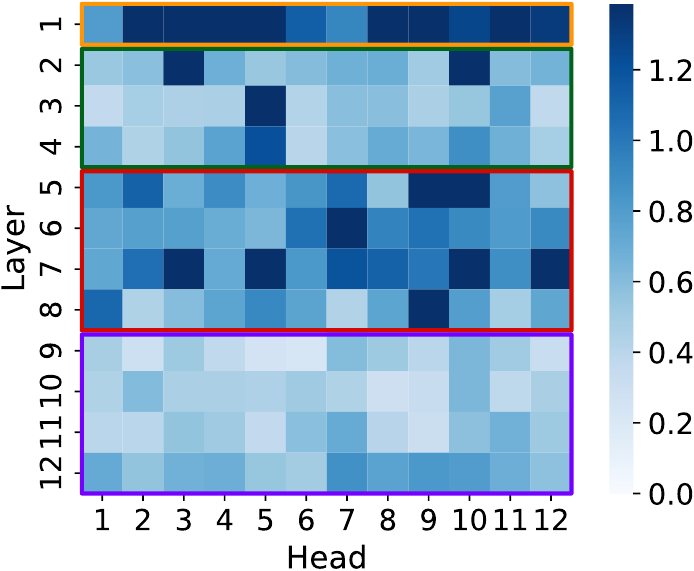}
        \end{subfigure}
    \caption{\clstok{}}   
    \end{subfigure}
    \hspace{-1.5cm}
    \begin{subfigure}[b]{.33\textwidth}
        \begin{subfigure}[b]{.15\textwidth}
        \includegraphics[scale=.26]{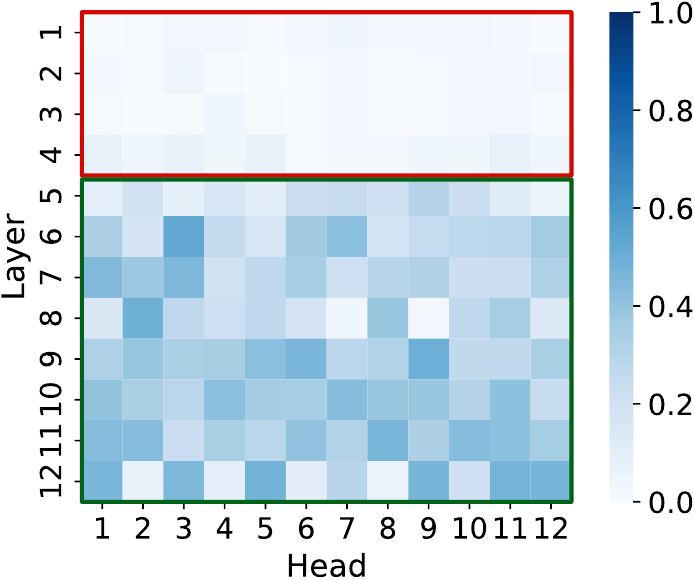}
        \end{subfigure}
        \hspace{2.1cm}
        \begin{subfigure}[b]{.15\textwidth}
        \includegraphics[scale=.26]{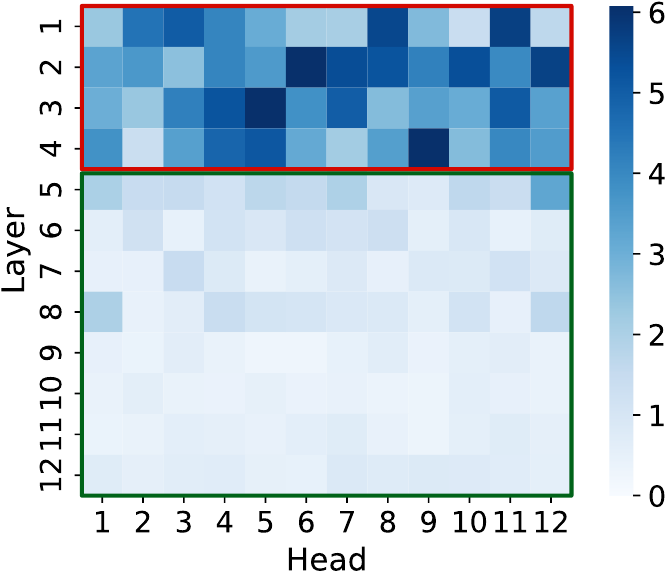}
        \end{subfigure}
    \caption{\septok{}}
    \end{subfigure}
    \hspace{-1.6cm}
    \begin{subfigure}[b]{.33\textwidth}
    \raggedleft
        \begin{subfigure}[b]{.15\textwidth}
        \includegraphics[scale=.26]{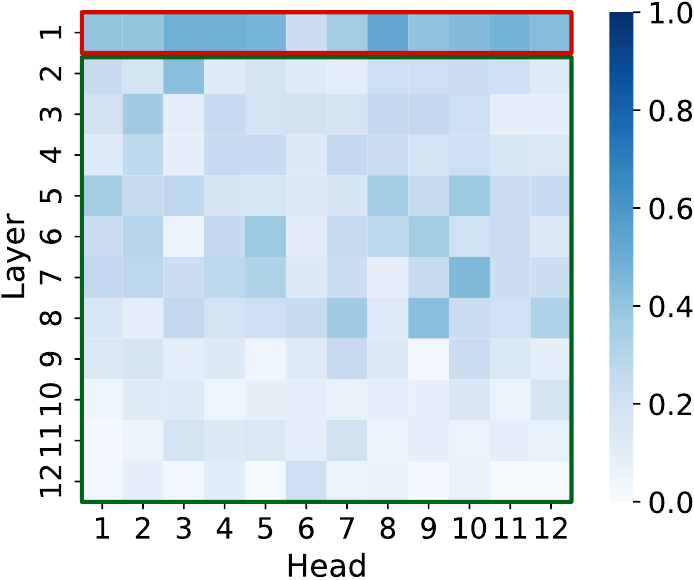}
        \end{subfigure}
        \hspace{2cm}
        \begin{subfigure}[b]{.15\textwidth}
        \includegraphics[scale=.26]{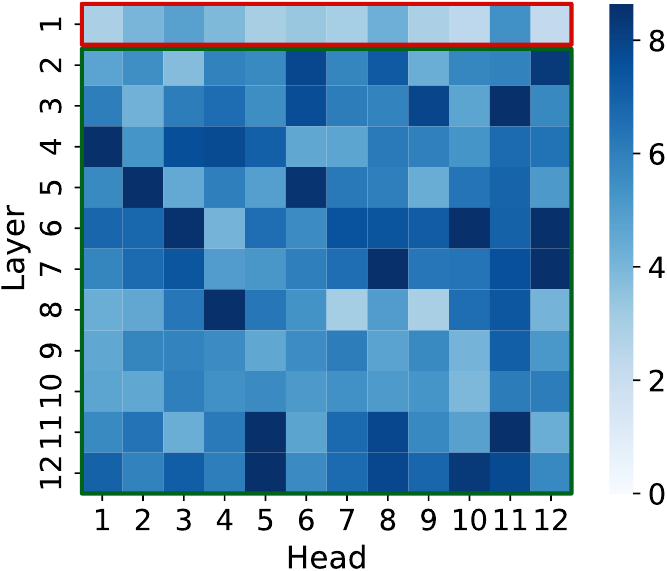}
        \end{subfigure}
    \caption{\textbf{\texttt{LTRL}}}
    \end{subfigure}
    \caption{Attention (\attn{}) and transformation norm maps ($||f(x)||$) for \clstok{}, \septok{}, and \textit{Literals} from the Java dataset. For each type of token, the left and right figures refer to the attention and norm map respectively. Regions that show the contrast relation are highlighted with the same colour.}
    \label{fig:alpha_vs_fx}
\end{figure*}

%% file: figures/ast_alpha_afx_agreement.tex
\begin{figure}[H]
    \begin{subfigure}[b]{.4\columnwidth}
    \hspace{-12mm}
        \includegraphics[scale=.33]{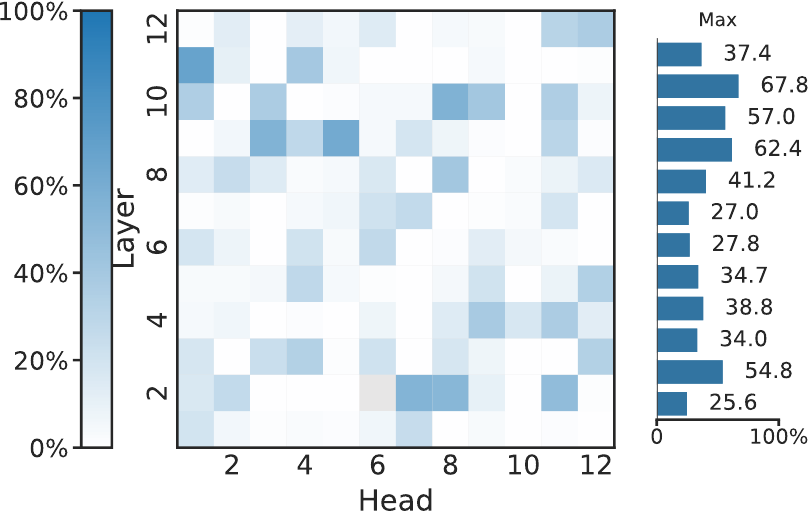}
            \caption{${p_\alpha}(g)$}
            \label{fig:ast_agreement_alpha}
    \end{subfigure}
    \hspace{-1mm}
    \begin{subfigure}[b]{.4\columnwidth}
    \raggedleft
        \includegraphics[scale=.33]{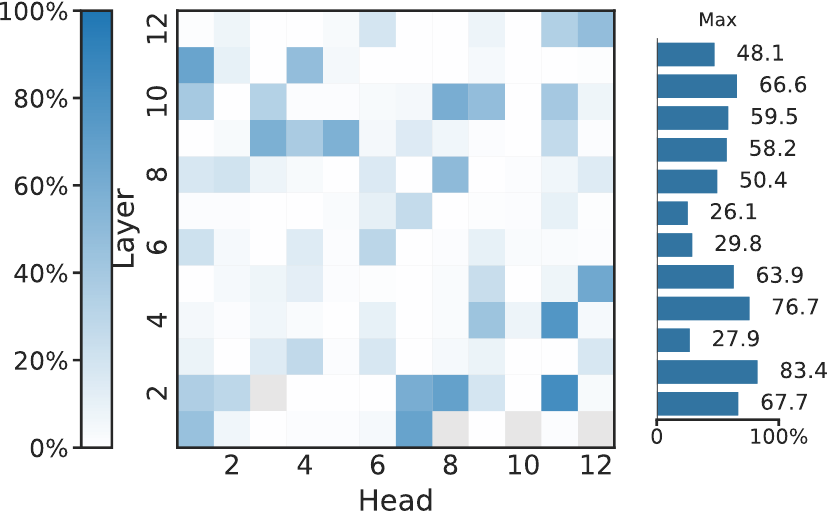}
            \caption{${p_{\alpha f(x)}}(g) $}
            \label{fig:ast_agreement_alpha}
    \end{subfigure}
    \caption{Agreement between the AST structure and the attention and scaled norm matrices on the Python dataset for all heads. We scale \afx{} maps within the $[0\ldots1]$ range. The bar plots at the right of each subfigure show the maximum agreement at each layer.} 
    \label{fig:ast_agreement}
\end{figure}

%% file: section/4_future_plans.tex
In this work, we have revisited the mathematical definition of \mhattn{} from prior works in natural language. We showed how the attention mechanism is not merely composed of the attention weights.
The preliminary findings indicate that incorporating scaled transformation norms provides new perspectives on what code properties are captured by attention.

The presented work can be extended in a variety of directions.
The first extension point is to investigate how these findings could vary across other programming languages and models. Applying the same methodology to models such as GraphCodeBERT~\cite{Guo2021} and CodeT5~\cite{Wang2021} and other languages such as JavaScript and Go will test the generalizability of our results.
Along similar lines, we aim to evaluate models trained with techniques that are more programming language-oriented. CodeBERT, despite being trained on NL-PL pairs, was pre-trained with the same objective as RoBERTa to model natural language. On the other hand, GraphCodeBERT encodes data flow paths in its input, which captures more source code properties other than its sequential nature. Similarly, CodeT5 uses an identifier-aware pre-training task that allows the model to determine which code tokens are identifiers and to recover them when they are masked. Determining if specialized training objectives modify attention behaviors will clarify how different representations are learned. The goal is to connect training procedures with resultant attention patterns.